 \documentclass[10pt,conference]{IEEEtran}
\usepackage[utf8]{inputenc}
\usepackage[T1]{fontenc}
\usepackage{amsmath}
\usepackage{amssymb}
\usepackage[]{graphicx}
\usepackage{paralist}
\usepackage{enumitem}
\usepackage{color}
\usepackage{xspace}
\usepackage[hyphens]{url}
\usepackage{hyperref}
\hypersetup{breaklinks=true}
\usepackage[numbers]{natbib}
\usepackage{csquotes}
\usepackage{mathrsfs}
\usepackage{afterpage}
\usepackage{rotating}
\usepackage{longtable}

\usepackage{multirow}
\usepackage[table,xcdraw]{xcolor}

\newif\iffull\fullfalse

\begin{document}
\title{A Blockchain based and GDPR-compliant design of a system for digital education certificates}
\author{
\IEEEauthorblockN{
    Fernanda Molina\IEEEauthorrefmark{1},
    Gustavo Betarte\IEEEauthorrefmark{1}\IEEEauthorrefmark{2} and
    Carlos Luna\IEEEauthorrefmark{1}\IEEEauthorrefmark{2}
    }
    \\
  \IEEEauthorblockA{
      \IEEEauthorrefmark{1}Área Informática, PEDECIBA}\\
  \IEEEauthorblockA{
    \IEEEauthorrefmark{2}Instituto de Computaci\'on,  
      Facultad de Ingeniería, Universidad de la Rep\'ublica\\
      Montevideo, Uruguay}\\
  Email: \{maria.molina, gustun, cluna\}@fing.edu.uy\\ 
  \IEEEauthorblockA{\vspace{5mm}} 
  \IEEEauthorblockA{} 
      }

\maketitle
\begin{abstract}
Blockchain is an incipient technology that offers many strengths compared to traditional systems, such as decentralization, transparency and traceability. However, if the technology is to be used for   processing personal data, complementary mechanisms must be identified that provide support for building systems that meet security and data protection requirements. We study the integration of \textit{off-chain} capabilities in blockchain-based solutions  moving data or computational operations outside the core blockchain network. We develop a thorough analysis of the European data protection regulation and discuss the weaknesses and strengths, regarding  the security and privacy requirements established by that regulation, of solutions built using blockchain technology   We also put forward a methodological framework that helps systems designers in combining operational off-chain constructs  with traditional blockchain functionalities in order to build more secure and privacy aware  solutions. We illustrate the use of that framework presenting and discussing the design of a system that provides services to handle, store and validate digital academic certificates.

\end{abstract}
\section{Introduction}
\label{introd}
Blockchain is an incipient technology that breaks some paradigms. It provides support to build decentralized systems where transactions are processed by the participating nodes of the network, without a responsible intermediary or authority. Blockchain offers many strengths compared to traditional systems, such as decentralization, transparency and traceability. On the other side, this technology has some general weaknesses concerning scalability and performance issues, but most importantly in our view, with confidentiality, immutability and access control. 

For this reason, if this technology is in particular to be used for processing personal data, complementary mechanisms must be identified that provide support to try to guarantee secure manipulation of that data. 
The approach we have followed to that respect is to study the incorporation in blockchain-based solutions of \textit{off-chain} capabilities, moving data or computational operations outside the core blockchain network.

Current proposals of off-chain processes aim to leverage a blockchain solution by addressing some of the intrinsic functional weaknesses described above. Typical scenarios that are identified as requiring the use of off-chain solutions are those that for their operation need to  improve performance or cost calculation processing, to perform intermediate operations on the off-chain leaving the final transaction on the blockchain (off-chain Signatures Pattern~\cite{10.1007/978-3-319-67262-5}) or to perform a final complex calculation on the off-chain (challenge Response Pattern~\cite{10.1007/978-3-319-67262-5}).  

One of the main objectives of our work has been to develop a methodological framework that helps system designers to select operational off-chain constructs that integrated with traditional blockchain functionalities allow to build more secure and privacy aware  solutions.
In particular, we have carried out a thorough analysis of the GDPR~\citep{GDPR-law} regulation in order to determine the weaknesses and strengths of solutions built using blockchain technology related to the security and privacy requirements established by that regulation.  Of special interest was to try to give a solution to  controversial subjects such as pseudonymisation and data anonymization when using hash functions and public key cryptography.

In the first phase of the investigation we have focused in developing two main constructs of the framework: a software architecture model and a use case model. The software architecture model embodies, among other components, an access control and audit network as well as an integrity network. The use case model consists of several use cases that cover the principal services we  understand can be used to build a blockchain and off-chain based system compliant with well established security and privacy requirements, in particular those established by the GDPR.

\subsection*{Related work and contributions}
There exist several proposals for off-chain solutions, the most common ones being either an external datastore, an external server or an external peer-to-peer network. In \cite{SOK} two differente off-chain models are proposed. The \textit{distributed or channels} model  consist of a group of equal nodes in a peer-to-peer network outside the blockchain, which are organized by pre-defined rules. The \textit{commit-chains or centralized}  model consists of a centralized system which receives and processes user requests and periodically responds to the chain.  The result of the procesing is transmitted to the blockchain, which in turn verifies the result before persisting it. In order to carry out that verification without disclosing confidential information the use of zero-knowledge test and verification processes like zk-SNARKs~\cite{zkSNARKs} and zk-STARK \cite{10.1109/Cybermatics_2018.2018.00199} has been proposed.
An off-chain solution may also be conceived as a storage or external processing system or as an hybrid one~\cite{10.1145/3284764.3284766}. In \cite{10.1007/978-3-319-67262-5} different off-chain processing patterns are proposed.  In this paper we put forward a consolidated understanding of those proposals providing guidelines to select off-chain models and their corresponding architectures and  technologies according to the problem to be solved.

Some of the challenges concerning data protection requirements the GDPR regulation poses to solutions built using blockchain technology are discussed in \cite{EPRS-art29} and \cite{EUBOF}. In those works it is analyzed the processing of pseudonymised and anonymized data and the potential privacy violations that might occur from the use of hash values and private/secret keys in blockchain solutions.  In \cite{DBLP:journals/corr/abs-1902-06359},  \cite{DBLP:journals/corr/abs-1904-03038} and \cite{10.1109/SC2.2018.00018}, different software architecture are proposed and discussed that have been conceived to make use of blockchain mechanisms to perform access control to data and auditing and to use off-chain solutions to safely storage and process  personal data. We propose to use additionally the blockchain as an integrity network in order to validate data stored off-chain, a link to data that is stored off-chain network is kept in the blockchain network so that it is possible to locate the data. Together with this information, it can be stored an integrity control check mechanism, for instance using a hash function.

In this paper we also put forward a methodological framework to specify the behavior and services of a system that integrates blockchain and off-chain functionalities. The two main components of the framework are a software architecture model and a use case model, which have been conceived to provide support for the construction of systems that are GDPR compliant by design. 

We illustrate the use of the proposed framework providing insights on  the implementation of a prototype of a system that provides services to handle, store and validate digital academic certificates.

\subsection*{Paper organization}
The rest of the article is structured as follows. 
Section~\ref{background} provides a primer on blockchain and off-chain concepts and models. In Section~\ref{gdpr} we analyze the European data protection regulation and discuss how well blockchain and off-chain mechanisms adapt to provide support for building  GDPR-compliant digital systems. The methodological framework is presented and discussed in Section~\ref{framework} and in Section~\ref{poc} we briefly discuss a proof of concept we are carrying on. We conclude and discuss further work in Section \ref{conclusion}. Finally, in the Appendix 
 we provide the complete specification of the use case model.


\section{A primer on blockchain and off-chain technologies}
\label{background}


\subsection{Blockchain basics}
\label{backg:blockchain}
Blockchain is a peer-to-peer system which builds a chain of blocks with no centralized authority. A blockchain network is composed of a set of transactions grouped in blocks. Each transaction is a unique cryptographically signed instruction that represents the valid passage from one state to another. A transaction can be a message or a code (usually a \textit{smart contract}), and can include a payment for its execution. Through a consensus protocol it is defined which node publishes the new block. Each time a new block is created, it is downloaded, processed and validated by all the nodes in the network. Thus, during that process each node executes all transactions contained in the block. Being a decentralized system, the nodes of the network have a copy of the entire chain and are responsible for validating and processing the blocks.

A smart contract is a program that is stored and executed in a blockchain without someone from the outside being able to interfere with its operation. For this reason, by design a smart contract is executed in a distributed way through consensus protocols and each node executes all the programs that are executed on the platform. Using smarts contracts it is possible to build a system where agreements are forced to be reached autonomously, since it is an algorithm that forces compliance. A relevant known example of smart contract network is the blockchain Ethereum~\cite{wood2014ethereum}.

Regarding the access control of users, blockchain networks can be categorized into two types: 
\begin{inparaenum}[i)]
 \item \textit{Permissionless,} the network is open to anyone willing to participate, so the level of distrust among the participants is high. Examples of these networks are Bitcoin~\citep{Bitcoin} and Ethereum; and 
 \item \textit{Permissioned,} the system is a private network where the entry of new participants is controlled, such as, for instance, Hyperledger~\cite{hyperledger}.
 \end{inparaenum} 
 
Blockchain security relies on public key cryptography for identifying transactions and hash technology to provide guarantees of the immutability of the chain: each block of the chain contains the hash value of the head of the previous block, therefore, a change in a block implies a change in the whole chain. This architecture ensures integrity, immutability and traceability by design.

However, since all transactions must be validated and processed by all the nodes of the network, all the information necessary to perform this processing must be public, what undermines the confidentiality of the information. The assets involved in the transactions are associated with private keys. Therefore the theft of those keys is a security risk any blockchain must endure. Digital assets might become unrecoverable in the case of theft of private keys, especially due to the lack of an administrator or system controller~\cite{deloitteInfo}. To mitigate this issue, it is essential to follow good practices in the use of wallets and key management. A risk related to transactions is the one of double spending, that is, spending the same amount of money more than once. This can happen when some consensus protocols are used that allow for multiple simultaneous chains to be generated over a period of time, until one is dropped \cite{Chohan2017TheDS}. Typically a blockchain network presents performance problems, since the core functional components of this technology, namely, transaction validation, consensus protocols and decentralization, require for all the transactions to be validated, processed and stored in all the nodes of the network.


\subsection{Off-chain concepts and models}
\label{backg:offchain}
In some circumstances it may be desirable to move data or computational operations outside the blockchain. This storage or external processing is called \textit{off-chain processing}. There exist several proposals for off-chain solutions, the most common ones being either an external datastore, an external server or an external peer-to-peer network. An off-chain solution may be conceived as a storage or external processing system or as a hybrid one~\cite{10.1145/3284764.3284766}. 

Storing large amounts of data could be expensive in blockchain, so a possible solution is to store the data outside the network, in a \textit{off-chain storage}, leaving a pointer on the network to the location of the data. To ensure the integrity of the data stored externally, a verification step should be performed. An option to validate the data stored in the chain is to keep in the blockchain a hash value of the data that is externally stored. In order to preserve the confidentiality  it is necessary to implement an external storage access control system. 

An alternative off-chain model is \textit{off-chain computation}, where a part of the processing is performed on the off-chain. The result of the execution is transmitted to the blockchain, which in turn verifies the result before persisting it. This verification is very important to ensure that the result returned is correct, since external processing should in itself be considered unreliable. 
Finally there are cases in which it may be necessary to use a \textit{hybrid model} ,using off-chain to store data and to perform computational process \cite{10.1145/3284764.3284766}.


From the operational and architectural point of view two different off-chain models have been proposed in \cite{SOK}. The \textit{distributed} or \textit{channels} model  consist of a group of equal nodes in a peer-to-peer network outside the blockchain, which are organized by pre-defined rules, for example, through a smart contract. This system requires using a peer-to-peer protocol between participating nodes, such as the Whisper Messaging Protocol \cite{Whisper}. In general these systems requires an unanimous consensus of the participants, since only transactions that are approved and signed by all nodes are considered valid. A node can at any time dump the approved off-chain calculations into the blockchain. In the same way, if there is a dispute about the outcome of the off-chain, such as a dishonest participant trying to lie about it, honest participants can resolve the dispute on the blockchain. 

On the other sidem the \textit{commit-chains} or \textit{centralized}  model consists of a centralized and not necessary realiable system which receives and processes user requests and periodically responds to the chain. This system performs the processing and returns the result along with a proof that it is correct, which is verified by the blockchain, using mechanisms of zero-knowledge test and verification process.

The incorporation of off-chain processes may introduce problems of integrity and availability of data. In a blockchain solution integrity is guaranteed by default, because by design it is impossible to modify data without being noticed by the rest of the network. However, when storing data outside the blockchain this feature is lost since a third party becomes responsible of handling the storage. Therefore, a verification process must be required to ensure the integrity of the stored information. One way to do this is to store in the blockchain a reference and a hash value  of the information stored externally, so that the information can be corroborated in case of alteration. Delegating the computational process also causes problems of integrity of the result, since the calculation is not performed by all the nodes of the blockchain, but it is instead performed by an external one. In a commit-chains model one way to ensure the correctness of the result obtained outside the blockchain is through the application of zero-knowledge test and verification processes. In channels models the dispute are resolved on the blockchain. 

As to availability, when data is stored outside the blockchain this property is difficult to guarantee since there is a single point of failure. Thus, solutions, , such as IPFS~\cite{IPFS} and SWARM~\cite{SWARM}, have been proposed where information is stored in a decentralized and redundant manner. Finally, off-chain solutions do not define by design audit processes, so auxiliary mechanisms must be considered in order to register an audit trail of the accesses and changes made.

In Table~ \ref{tab:Framework to select offchain models} we present guidelines to select off-chain models and their corresponding architectures and technologies according to the problem to be solved

\begin{table*}
\renewcommand{\arraystretch}{1.3}

\caption{Off-chain models and implementation}
\label{tab:Framework to select offchain models}
\centering
\begin{tabular} { |  p{4.5cm} | p{5cm}| p{7.2cm}|  } 

    \hline
     \textbf{Problem to solve} & \textbf{Model }& \textbf{Technology suggested} \\
    \hline
    \hline
    
 \textbf{Improve calculation processing}  & Off-chain computation with channels model & Whisper Messaging Protocol, Zero-knowledge verification process (Zk-SNARKs or zk-STARKs) \\
 \hline
\textbf{Improve processing of a final calculation} & Off-chain computation with channels model &  Whisper Messaging Protocol \\
\hline
 \textbf{Reduce calculation processing costs} & Off-chain computation with channels model & Whisper Messaging Protocol, Zero-knowledge verification process (Zk-SNARKs or zk-STARKs) \\
\hline
\textbf{Ensure the cost of intermediate transactions} & Off-chain computation with channels model. Send final transaction to blockchain & Whisper Messaging Protocol \\
\hline
\textbf{Reduce storage costs} & Off-chain storage with channels or commit-chain model  & Hash verification integrity \\
\hline
\textbf{Ensure the confidentiality of data required to perform a calculation} & Off-chain computation with commit-chains model & Zero-knowledge verification process (Zk-SNARKs or zk-STARKs) \\
\hline
\textbf{Ensure the confidentiality of the information to be stored} & Off-chain storage with commit-chain model & Hash verification integrity, external storage access control system \\
\hline
\textbf{Ensure the confidentiality and availability of information to be stored} & Off-chain storage with commit-chain model & Hash verification integrity, external storage access control, SWARM, Interplanetary File System \\
\hline
\end{tabular}
\end{table*}

\section{Blockchain constructs, data protection principles and requirements} 
\label{gdpr}
The GDPR \cite{GDPR-law} is the European regulation on the protection of natural persons with regard to the processing of personal data and the free circulation of these data. The GDPR entered into force on May 25, 2016 and came into effect on May 25, 2018. During that time interval companies and organizations were required to adapt to comply with that law. European countries had their own personal data protection laws. With the GDPR, they became governed by a common legislation, which not only reaches companies or organizations resident in countries belonging to the European Union, but also foreign companies or organizations that deal with data of EU residents. 


\subsection{Scope, roles and responsibilities}
The GDPR defines in its \textit{Article 4} that personal data includes all the data that is or can be assigned to a natural person, such as, for instance, the phone number, credit cards, account information, registration numbers, appearance, customer number or address. As discussed in \cite {EPRS}, the reference to an identifiable person indicates that it is not required the data to be identified as belonging to someone to qualify as personal data, but that the mere possibility of identification is sufficient. This concept is important in the context of blockchain solutions, where individuals can be identified through the use of public keys. 

In its \textit{Article 4}  the GDPR defined the roles that are responsible for the handling of personal data:
\begin{inparaenum}[i)]
 \item \textit{Data Controller}: It is responsible for the processing of information and the appointment of the processor role (\textit{Article 28}). A data controller can process the data collected using its own processes, or it can also work with a third party or an external service to process the data that has been collected. Even in this situation, the data controller will not transfer control of the data to the third-party service, as it will be responsible for specifying how the external services will use and process the data,
 \item the \textit{Data Processor}: a data processor processes the data that is provided to him by the data controller, but he does not own the data he processes or controls. This means that the data processor cannot change the purpose and the means in which the data is used, since this is defined by the data controller,
 \item the \textit{Representative}: \textit{"A natural or legal person established in the Union who, designated by the controller or processor in writing pursuant to Article 27, represents the controller or processor with regard to their respective obligations under this Regulation"}, 
 \item the \textit{Owner}: is the owner of personal information, and finally
 \item  the \textit{Recipient}: a natural or legal person, public authority, agency or another body, to which the personal data are disclosed, whether a third party or not.
 \end{inparaenum}

\subsection{Pseudonymisation and data anonymization}

The GDPR differentiates the processing of pseudonymised and anonymized data. With regard to pseudonymised data, the \textit{Recital 26 of the GDPR} states that this type of information is under the scope of the GDPR, while the principles of data protection should therefore not apply to anonymous information. Pseudonymisation is important for risk minimization (\textit{Recital 28 of the GDPR}), but it should not be seen as an anonymization technique. Related to this, the document \cite {EPRS} analyzes the \textit{Article 29 Working Group} (an independent European working group that has dealt with issues related to the protection of privacy and personal data) \cite {EPRS-art29}, who defines some pseudonymization techniques used by blockchain, such as encryption with secret keys and hash functions. They are considered pseudonymization techniques because it is still possible to obtain the original data which is supposed to be protected.


\subsection{Personal data management using blockchain}
The GDPR guarantees rights to the owner of personal data that is managed by a third party, including the right to access the data and the right to erase and rectify it. The Law also requires data protection and transparency in the processing of that data. In what follows we go through those rights analyzing how compliant is the management data that resides in a system built using blockchain mechanisms.
\subsubsection*{The right of access to personal data}
In GDPR, \textit{Articles 12 to 15 and Article 20 of the right to data portability}, indicate the right to request information about the data and the obligation to provide it within a month. That information might concern the identification of the stored data, to whom it has been transmitted, the period of retention and the existence of automated decisions on the data, among others. This type of access to data is facilitated with blockchain, since the data is, in fact, available to all members of the network.
\subsubsection*{Confidentiality of personal data} 
Confidentiality can be a weak point in blockchain, since the data resides by design in all the participating nodes of the network. That is the reason why, if the necessary measures are not implemented, all personal data can become accessible to all people with access to the chain.
\subsubsection*{Deletion of personal data} 
The European Law considers the obligation to erase or modify the data at the request of the owner of the personal data (\textit{Article 16 - Right of rectification}; \textit{Article 17 - Right to be forgotten}; and \textit{Article 18 - Right to limit the treatment of the GDPR}), which implies the need to delete data, either when the data is not correct or the consent of the owner of the data is withdrawn, or when the stipulated period of use or the purpose for which the data was used ends. In turn, the data must be stored for a specific purpose and deleted when the service is finished (\textit{Article 23 - Limitations of the GDPR}). In blockchain a stored data cannot be modified or deleted once it is in the chain, so alternatives must be sought to meet this requirement, such as storing personal data outside the blockchain, for instance in a off-chain storage, and saving a reference to the blockchain. As to the right of rectification, it is impossible to modify the data in a block once it resides in the blockchain, so one way of achieving this is by entering the updated data in a new block and allowing a subsequent transaction to cancel the initial transaction, even if the first transaction is still in the chain.
\subsubsection*{Traceability of personal data} 
The GDPR Law states (\textit{Article 30 - Registration of treatment activities}) the need to keep a trace of the activity carried out with personal information, such as to whom has the data been shared. Additionally, \textit{Article 15 - Right of access of the interested party}, indicates the right of the data owner to receive information regarding the processing of his data, including with whom has the information been shared. The activity log is natural in blockchain, where everything is stored in the chain and all participants have the information of the system transactions, so these requirements are met by design. 
\subsubsection*{Responsibility for data processing} 
Regarding the responsibility in the processing of personal data in a blockchain network, the cases of permissionless and permissioned networks should be considered separately. In the case of permissioned networks, it is easier to define who is the responsible, so whenever possible it is better to use this type of networks. In an permissionless blockchain network this role cannot be assigned to a person or institution, but it can be analyzed if it is admissible in the legal field to consider the shared responsibility of all or some of the members of the blockchain. This is considered in the GDPR in \textit{Article 26}, as \textit{Joint controllers for the treatment}. In a blockchain network we can identify several actors, such as network developers, nodes, miners or smart contract developers. However, there is no common consensus on which actors have responsibility for data processing and what is clear is that we are in a grey area where, in some cases, it will not be possible to identify a controller. 
 

\subsection{Public/private keys as identifying data}
Blockchain uses a public/private key system to identify the owners and recipients of the transactions, so it is quite crucial to analyze how this information is managed and whether it is considered or not as personal data. 

Related to this, \textit{Recital 30 of the GDPR } states that persons may be associated with online identifiers and, as discussed in \cite {EPRS}, in the context of blockchain, public/private keys serve as the type of identifiers mentioned in this recital, since they are often used to identify the origin and destination of each transaction and to sign the transactions. Therefore,  they should be treated as personal data, that is, as an identifier of a person.


The problem is that in blockchain public keys cannot be obfuscated or deleted since they are used to identify transactions. Related to this, the document on blockchain and the GDPR of the \textit{European Union Blockchain Observatory and Forum}~\cite {EUBOF} indicates that one way of trying to avoid traceability is to use a new pair of keys for each transaction to prevent them from being linked to a common owner. The objective is to mitigate the risk, if the owner of a key is revealed, that the binding could reveal all other transactions that belonged to the same owner. 


\subsection{Hashes as pseudonymized information}

It can be inferred that applying a hash algorithm on personal data is a way of pseudonymizing that data, since it can indirectly identify a personal data. Thus, hash values used in blockchain must be considered personal data. Related to this the work reported in 
\cite{EUBOF} warns that when using a hash function, it is necessary to be aware that patterns that could allow traceability are not being created. This is due to the risk of bonding, what refers to situations in which pattern analysis allows an analyzer to discover information about a particular individual. The risk of bonding increases if simple information is saved, because the outputs of a hash function can be guessed from known inputs. Therefore, it is advisable to use a salt in the hash function as a means to reduce the probability of obtaining the input value. Nevertheless, the \textit{Article 29 Working Group}~\cite{EPRS-art29} makes it clear that the use of this technique does not produce anonymous data, since it is still possible to calculate the original attribute hidden behind the hash value. 

The \textit{Data Protection Agency of Spain} analyzes in \cite{AEPD} the reversibility of a hash function and concludes that, when the data used in a processing operation has an implicit order, the set of possible messages is reduced, which makes message reversal easier. Given a dataset to which a hash is to be applied, the degree of entropy of the dataset influences the reversibility of the hash. The smaller the message space  the lower the entropy are, the lower the risk of collision in hash processing is, but re-identification will be more likely. On the contrary, the higher the entropy is the higher the possibility of a collision, but the risk of re-identification will be lower. In  \cite{AEPD} it is also stipulated that it is also necessary to take into account the identifiers linked to a hash value, because the more personal information is linked the higher the risk of identifying the contents of this hash value. Finally, is is also suggested using a single-use salt model that generates a separate random element for each message. This random element must be completely independent of any message and any other salt generated for any other message. 

Given the immutability of the data stored in blockchain, it is important to determine what can be considered acceptable for Personal Data Protection regarding data erasing. An alternative to erase the information is to process the data, instead of deleting it, so that it becomes anonymized and it is no longer within the scope of the GDPR. In the case of hashes, this would amount to eliminate the original data from which the hash was formed. Regarding anonymization techniques,  the work \cite {EPRS} concludes that the GDPR takes a risk-based approach, as it takes into account not only current technology, but also future one. \textit{Recital 26} of the GDPR states that for anonymization techniques it should be considered which is the available technology at the time of the processing. In that respect, the Article 29 Working Group \cite{EPRS-art29} postulates that the possible advancement of technology should be considered in the period of time in which personal data will be stored. In the case of blockchain this period of time is undefined, so in \cite{EPRS} it is argued that any data should be considered as personal data, since it cannot reasonably be assumed that identification will remain anonymous in the future. Despite these arguments, there are antecedents where anonymization has been accepted as a form of erasure. In fact, the \textit{Commission Nationale de l'Informatique et des Libertés (CNIL)}, in an article on blockchain~\cite{CNIL}, accepts the deletion of an original data as a method of deletion, even if the hash remains in the blockchain.


\subsection{Weaknesses and strengths of blockchain and off-chain solutions}
Considering the previous discussion, we understand that the data protection requirements defined by the GDPR can be hardly addressed alone by the blockchain technology and the concept of a ledger stored by all participants of the network. In permissionles blockchain networks, in particular, as there is no identified and assigned managers of the network it is not possible to ensure that access to the data is controlled and regulated. In the general case, we find it difficult for systems built using blockchain technology to guarantee the confidentiality, deletion or modification of the data managed in those systems. 

On the other side, blockchain technology meets by design some of the requirements of these Data Protection Laws, such as the right of access to data by the owners, traceability, encryption and hash techniques as security by design mechanisms and finally transparency in data processing. Furthermore, we have already argued that extending a blockchain solution with off-chain processing the requirements concerning responsibility for data processing, access control and the possibility to delete or modify personal data could also be satisfied.

\section{Privacy-oriented blockchain design guidelines} 
\label{framework}

According to the analysis presented in Section~\ref{gdpr}, it seems quite evident that the design, and implementation, of systems that use blockchain technology must incorporate complementary mechanisms in order to comply with the data protection requirements stipulated by the GDPR. 

In particular, our proposal is to consider an hybrid model in which off-chain is used to store personal data and blockchain to ensure integrity, traceability and access control. In short, personal data should not be stored on the blockchain, but instead register a link to be able to access the data residing in an off-chain storage. Additionally, it must be carefully evaluated what type of information is hashed and how this hash is done (using some salt, for example). In case of storing hashes in the blockchain, the possibility of anonymization as a deletion method should be argued, as described before. As an option, alternative methods that do not involve storing personal hashed information, such as using zero-knowledge generation mechanisms, should be considered. However, these mechanisms are rather intended to confirm that a third party has certain information or meets a certain requirement, so it is not always aligned with the needs of the business (such as verifying the integrity of stored data).
We also postulate that if a system is intended to use off-chain processing of personal data, it is preferable not to use channels model, since  the controversies in it are resolved by executing the processing in the blockchain, exposing personal information. A centralized model may be more appropriate. 

We put forward the following guidelines for the design of a system that integrates blockchain technology with off-chain processing of data and that is intended to comply with personal data protection requirements like those stated in the GDPR:
\begin{inparaenum}[i)]
\item In the design stage evaluate the impact related to data protection;
\item The treatment of smart contract data should be regulated, in accordance with what was agreed with the owner of the data;
\item Implement an access control system, so that only authorized persons can access the data;
\item People who access the data must sign a confidentiality agreement;
\item Consider implementing via smart contracts the consent of the owner of the data for its use;
\item Consider implementing the access of the owner of the data to his data, upon request;
\item Use encryption or hash functions as a means of data protection;
\item Transparency in the processing of information could be fulfilled by publishing the smart contract used;
\item Operations must be registered, such as to whom data was given and all the activity carried out;
\item Automatic deletion of data: when the consent of the owner of the data is revoked or the service has ended;
\item Integrity: implement evidence generation mechanisms such as hash values or zero-knowledge, to ensure integrity in the processing of information; and 
\item Availability: evaluate alternatives and which off-chain models to use to ensure the availability of the data.
 \end{inparaenum}

In what follows we introduce a methodological framework to specify the behavior and services of a system that integrates blockchain and off-chain functionalities.  The two main components of the framework are a software architecture model and a use case model, which have been conceived to provide support for the construction of systems that are GDPR compliant by design.

\subsection{A software architecture for hybrid blockchain and off-chain solutions}

We first proceed to schematically present in Table~\ref{tab:Blockchain and offchain complain GDPR} what are the blockchain and off-chain  functional components that combined make it possible to design a solution satisfying data protection requirements. In particular,  given a data security or privacy property we point out the behavior that can be (or can not be) guaranteed by a blockchain and off-chain mechanism. Associated to the property we also make explicit the design principles that shall constitute the methodological reference to build the architecture of hybrid models.

\begin{table*}[t]
\renewcommand{\arraystretch}{1.3}

\caption{GDPR-complying design principles}
\label{tab:Blockchain and offchain complain GDPR}
\centering
\begin{tabular} { |  p{2.5cm} | p{4cm}| p{4cm}| p{6cm}| } 

 \hline
\textbf{Point to comply} & \textbf{Blockchain} & \textbf{Off chain} & \textbf{Framework Considerations} \\
  \hline
  \hline

  \textbf{Responsibility} & If possible, use permissioned networks & Use storage off-chain with commit-chain model & Responsibility is based on private blockchain models and centralized off-chain models \\ 
 \hline

\textbf{Confidentiality} & Difficult to ensure the confidentiality of the data stored & Use commit-chain model & The data stored in off-chain must be encrypted, as well as the communications\\ 
 \hline
 
 \textbf{Access Control} & The access control policy must be stored in the blockchain &  Validate access control policy to authorize access to data & Offchain must validate authentication and authorization against the stored access policy \\
  \hline
  
\textbf{The right of access to personal data} & Solved by design in blockchain & Consider implementing the access of the owner to his own data, upon request &  Offchain must perform validate authentication and authorization using the  access policy stored in the blockchain \\
    \hline
    
\textbf{Delete/Modification of personal data} & The immutability of blockchain prevents to perform these operations &  Storage personal data in off-chain  & The process of deletion or modification must be authorized, as well as the process of access to data\\  
\hline

\textbf{Integrity} & Solved by design in blockchain & Additional integrity control mechanism must be implemented &  Store in blockchain a hash of the data stored in off-chain as an integrity control check \\
  \hline
  
\textbf{Traceability} &  Solved by design in blockchain & Offchain does not offer traceability by design & Use blockchain as audit log \\
  \hline
  
    \textbf{Transparency} & Smart contracts used in blockchain networks can be published & Minimize process in off-chain & All access to data on the off-chain must be recorded in the audit log, which is kept in the  blockchain \\
  \hline
  
\textbf{Availability} & Solved by design in blockchain & Issue to resolve when choosing commit-chain architecture & This point should be evaluated when instantiating an off-chain solution\\
  \hline
  
\end{tabular}
\end{table*}


Given the functional characteristics of the blockchain technology, it has been suggested, for instance in \cite{DBLP:journals/corr/abs-1904-03038}, to use those functionalities to build embedded in the system
\begin{inparaenum}[i)]
\item  an \textit{access control network}, that stores the access control policy and, upon request, performs the corresponding authentication and authorization tasks, and
\item an \textit{audit network} that records all actions that have been performed, such as authorizations, access requests, modification and deletion of data. It must be registered, for instance, who accessed the data and both approved and denied accesses. The immutability of blockchain provides an appropriate setting  for recording audit logs.
\end{inparaenum}

In addition to those two networks, we propose to use additionally the blockchain as an \textit{integrity network} in order to validate data stored off-chain. When a data is stored in the off-chain network, a link can be stored in the blockchain network so that it is possible to locate the data. Together with this information, it can be stored an integrity control check mechanism, for instance using a hash function. 

In this architecture, we can use a Gateway for external communication towards the blockchain and off-chain, similar to how is done in the architecture proposed in \cite{10.1109/SC2.2018.00018}. As authorization mechanism we propose to use a Security Token Service, like the one proposed in \cite{DBLP:journals/corr/abs-1904-03038}. In order for a user to access personal data, which is stored in the off-chain, first he has to obtain an authorization token from the access control network. When attempting to access the data, the user presents that token to the off-chain, which in turn validates it with the access control network.

\subsection{The use case model}
This model consists of eight use cases that cover the principal services we understand can be used to build a blockchain and off-chain based system compliant with well established security and privacy requirements.
It follows a brief characterization of each use case:

\begin{inparaenum}[i)]
\item \textit{Register personal data}, data is stored in the off-chain and registered in the blockchain;
\item \textit{Grant access}, authorization of access to personal data to a third party;
\item \textit{Revoke access}, removal of third party access permission;
\item \textit{Data access}, authenticated and authorized access to data stored in the off-chain;
\item \textit{Verify data}, integrity verification of data stored in the off-chain;
\item \textit{Delete or modify data from owner}, request of modification or deletion of data by its owner;
\item \textit{Delete or modify data from data controller}, request of modification or deletion of data controller; and
\item \textit{Request for access log}, requested by the data owner or an external authority.
 \end{inparaenum}

In what follows we shall describe in detail the use cases \textit{Data access} and \textit{Verify data}. The complete specification of the use cases are described in the Appendix.

As requestors of services provided by and to the system we consider the following actors:

\begin{itemize}
\item \textit{Data owner (DO)}: the user, owner of personal data;
\item \textit{Data controller (DC)}: responsible for data processing;
\item \textit{Data processor (DP)}: a data controller can delegate processing tasks to a data processor; and
\item \textit{Receiver (recipient)}: third party who wants to access personal data.
\end{itemize}

These actors must be registered in the system to be able to perform authentication and authorization operations. For this, different solutions can be used, for example using a system of certificates with private-public key to authenticate data owner and a user and password system for recipients.

\subsubsection{Data access use case}

\begin{figure} [h]
 \centering
    \graphicspath{{./images/}}
    \includegraphics [width = 0.35 \textwidth] {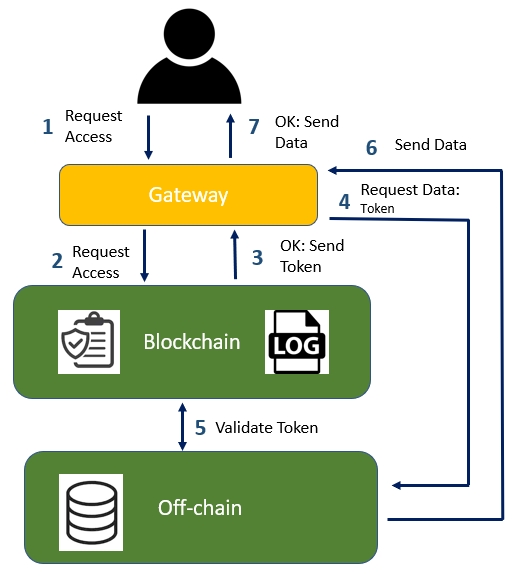}
   \caption{Use case diagram: Data Access}
   \label{fig: UCD - Data Access}
\end{figure}

\begin{figure} [h]
 \centering
   \graphicspath{{./images/}}
    \includegraphics [width = 0.48 \textwidth] {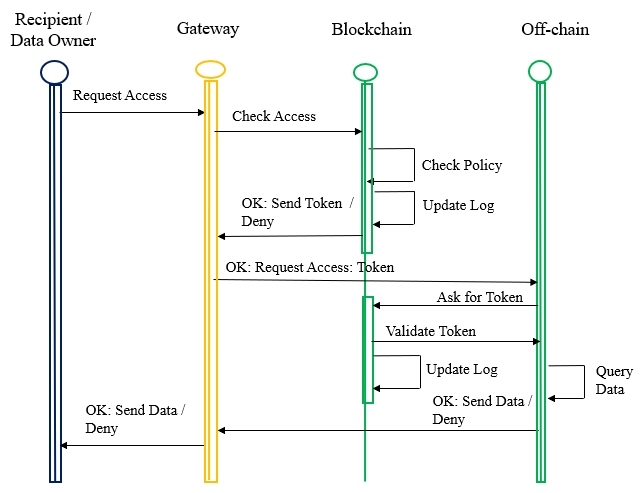}
   \caption{Sequence diagram: Data Access}
   \label{fig: SD - Data Access}
\end{figure}

When a data owner or recipient wants to access to personal data, it has to send a request to the Gateway. The Gateway then sends the request to the access control blockchain network, which validates the access by reviewing the defined policy and, if the person requesting access is authorized, the blockchain delivers to the Gateway a token and the link to the data stored in the off-chain network. With this information, the Gateway can request access to the off-chain network, which in turn validates the token against the access control blockchain. If the Token is validated, the off-chain network returns the data and the Gateway sends the data to the requestor. All these accesses are stored in turn in the blockchain audit network. The use case diagram is illustrated in Fig. \ref{fig: UCD - Data Access} and the control and data flow just described is formally specified by the sequence diagram illustrated in Fig. \ref{fig: SD - Data Access}.

\subsubsection{Verify data use case}
\begin{figure} [h]
 \centering
    \graphicspath{{./images/}}
    \includegraphics [width = 0.35 \textwidth] {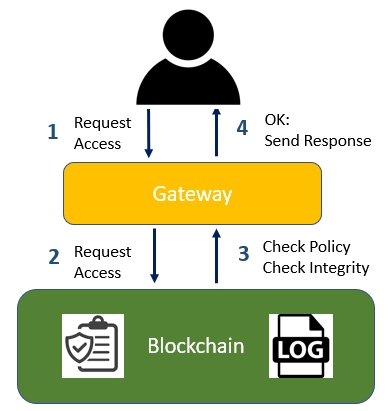}
   \caption{Use case diagram: Verify Data}
   \label{fig: UCD - Verify Data}
\end{figure}

\begin{figure} [h]
 \centering
    \graphicspath{{./images/}}
    \includegraphics [width = 0.48 \textwidth] {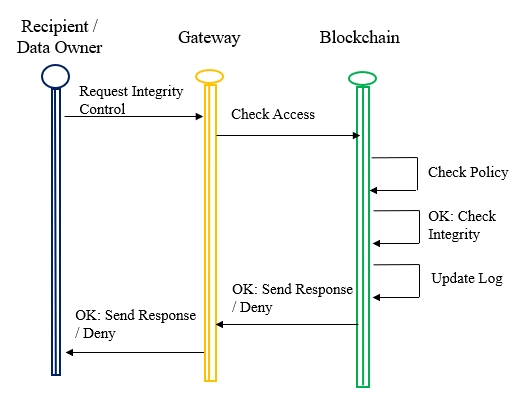}
   \caption{Sequence diagram: Verify Data}
   \label{fig: SD - Verify Data}
\end{figure}

Once personal data is available in the system, the data owner or a third party might be interested in verify its integrity. We take the view that the processes of validating personal data and accessing the hash used to perform the validation are constitute sensible operations. Therefore, it is necessary to implement access control for this use case, as is represented in Fig. \ref{fig: UCD - Verify Data}. To do that, the Gateway first validates access against the access control network blockchain. Once authorized, the integrity blockchain performs the integrity control of the data passed by the user. This is done by comparing the stored hash with the one presented by the user. It should be noted that in this use case it is not necessary to consult the off-chain. This control and data flow is formally specified by the sequence diagram described in Fig. \ref{fig: SD - Verify Data}.

\subsection{Some implementation considerations}
In addition to the design constructs just described, it is also important to evaluate which network models will be adopted, as well as other implementation considerations. 

As previously discussed in Sect. \ref{gdpr}, the creation of a permissioned blockchain network is recommended to handle personal data. In turn, the use of an off-chain with commit-chain architecture network is recommended, but, being a centralized and non-distributed system, this implies having to design solutions to ensure system availability. 

As already mentioned, a hash value of data could be stored in the blockchain as an integrity control mechanism. Thus, the proposed validation is based on the comparison of the hash of the data stored in the blockchain with the data stored in the off-chain. To do this safely, namely avoiding the hash reversibility problem, it is necessary to consider the recommendations given in Sect. \ref{gdpr}, designed to increase the entropy level of possible hashes. That implies, for instance, introducing random elements and taking into account the identifiers linked to a hash. Related to this, it is important to carefully choose the certificates and public keys to be used to identify the actors of the system, for example, analyzing if it is applicable to use a new pair of keys for each transaction in order to a void traceability.

\section{Applying the framework: a case study on digital titles} 
\label{poc}

\subsection{The problem}
We are currently working in the implementation of a prototype of a system that provides functionalities to handle, store and validate university certificates, like degree diplomas. The IT services unit of the Universidad de la República, Uruguay, which is called SeCIU (Servicio Central de Informática) is in the process of studying alternatives for implementing and deploying one such system using blockchain technology. The main objective of the solution is that (graduated) students are able to obtain their digital titles or certificates and deliver them to third parties (for example potential employers), which in turn could be able to validate the authenticity and legitimacy of those certificates using that same system.

Institutions all around the word have been concerned for a long time with finding good solutions to solve the problems related to educational certificates. There exists worldwide a severe problem of fake degrees, which is aggravated by the fact that, for instance, employers do not have the ability to validate the certificates a candidate present when applying to a job post. In the context of certification of studies, the blockchain technology has been visualized as the solution to counter the manipulation of fake certificates by providing an easy validation mechanism that provides integrity assurance.


\subsection{Analysis of existing solutions}

There exist several solutions that have been proposed to deal with this problem, with different degrees of development. In particular, SeCIU has analyzed Latin American solutions like the Brazilian RAP System and the Argentine System BFA. In \cite{FakeDip} the solutions \textit{Blockchain for Education} and \textit{EduCTX} are discussed. All these systems propose a hybrid solution, using the blockchain to perform the validation of the certificates by storing a hash of the certificate and, in some cases, some additional information. Some of those  solutions use public blockchain networks and others private ones, but none of them implements access control mechanisms to perform the certificate verification process.
For the execution of that process, the candidate gives the employer his certificate (an act which is considered an implicit consent of personal data access) and the employer validates the provided certificate with the blockchain, typically comparing hash values. However, as previously pointed out in Section \ref{framework}, performing this verification implies access to personal data and therefore the execution of that procedure should be authorized by the data owner.

We now proceed to describe in more detailed the solutions we have analyzed:

\begin{itemize}
\item \textit{Brazilian RAP System} \cite{RAP}. In Brazil, there exists a regulation that sets a 2-year period from March 2019 for universities to implement a diploma verification system. Universities are working to generate digital diplomas, using signed XML documents and blockchain. The RAP system propose to use the public networks Ethereum and Bitcoin to provide existence and integrity controls, and off-chain network to preserve digital documents. Additionally there is an authentication module, that performs data validation with the blockchain network and retrieves the information from the off-chain network. 
\item \textit{Argentine BFA System}, which was developed by the University of Córdoba and based on the BFA (Blockchain Federal Argentina) \cite{BFA}. BFA is based on Ethereum technology and works under the model of a permissioned blockchain. Once the University records are validated, a digital document is saved and digitally signed by the teacher, and then is stored on the blockchain. In BFA no documents or files are stored within the blockchain, only the hashes of those documents are saved. 
\item \textit{Blockchain for Education} \cite{BFE}, consists of a network of universities that have implemented a hierarchy system to add Universities to the system and to validate the signing certificates. There is an accreditation authority, responsible for authorizing the entry of other universities to the system and certification authorities, that are the universities that belong to the system, responsibly for signing certificates and storage it in the blockchain. BFE uses a public network Ethereum so, like adding any transaction on this network, adding certificates comes into a cost, that is a disadvantage of the system. It also uses a smart contract for identity management and another smart contract to manage and store certificates in the blockchain. 

\item \textit{EduCTX} \cite{EDUCTX} proposes a global higher education credit platform, based on the concept of the European Credit Transfer and Accumulation System (ECTS). It constitutes a globally trusted, decentralized higher education credit and grading system, based on a permissioned blockchain network of higher education institutions (HEI). In this system, every time a student completes a course or save an exam, his HEI will transfer the appropriate number of ECTX tokens to his blockchain address. When an organization wants to verify the student's course obligation completion, the student has to send his blockchain address and redeem script to the verifier organization. Then the organization checks the amount of ECTX tokens againt the blockchain, which represents the student's academic credit achievements.
\end{itemize}



\subsection{The design of the solution}
We have specified the expected behavior of the system using the GDPR-compliant constructs of the framework  introduced in Section~\ref{framework}. In what follows we shall present and discuss the most relevant design decision we have taken. 

In the first place, we have defined the following mapping from the data related roles presented in Section~\ref{framework} to the institutions and individuals pertaining to the problem's domain:
\begin{inparaenum}[i)]
\item \textit {SeCIU} is the \textit {DC}, the responsible for the personal data handled by the system; 
\item the \textit {School Registry Offices} are the \textit {DP}, working as delegates assigned by the SeCIU; 
\item the \textit {Candidates} are the \textit {DO},  the owners of personal data;
\item and finally the \textit {Receiver (recipient)} are the employers  or institutions that want to validate a certificate.
\end{inparaenum}

Then, degrees and schooling (that corresponds to personal data) are stored in an off-chain network under the responsibility of the DC, and the operations of audit, access control and verification are carried out in a blockchain network that is accessed by School Registry Offices, students and authorized third parties, through a Gateway.

To validate the certificates, a hash of each certificate is stored in the blockchain, so that the validation system is based on the comparison of this hash with the hash of the certificate presented by the student. As was discussed in Section~\ref{gdpr}, hash values should be considered personal data as they are pseudo-anonymized data, then access to that information should be authorized by the DO. Therefore, certificate verification should not be a public function and access to that operation should be controlled. In this respect, the authorities of the Universidad de la República decided to formally ask the Uruguayan Data Protection Agency, the \textit{URCDP} (\textit{Unidad Reguladora y de Control de Datos Personales}), whether it is valid, from the legal point of view, to implement a system where the verification of a degree diploma is public. The response of the URCDP was that such verification requires the consent of the individual that earned the diploma  and therefore cannot be provided as an open and public function~\cite{URCDP}. 

\begin{figure*} [h]
 \centering
    \graphicspath{{./images/}}
    \includegraphics [width = 0.75 \textwidth] {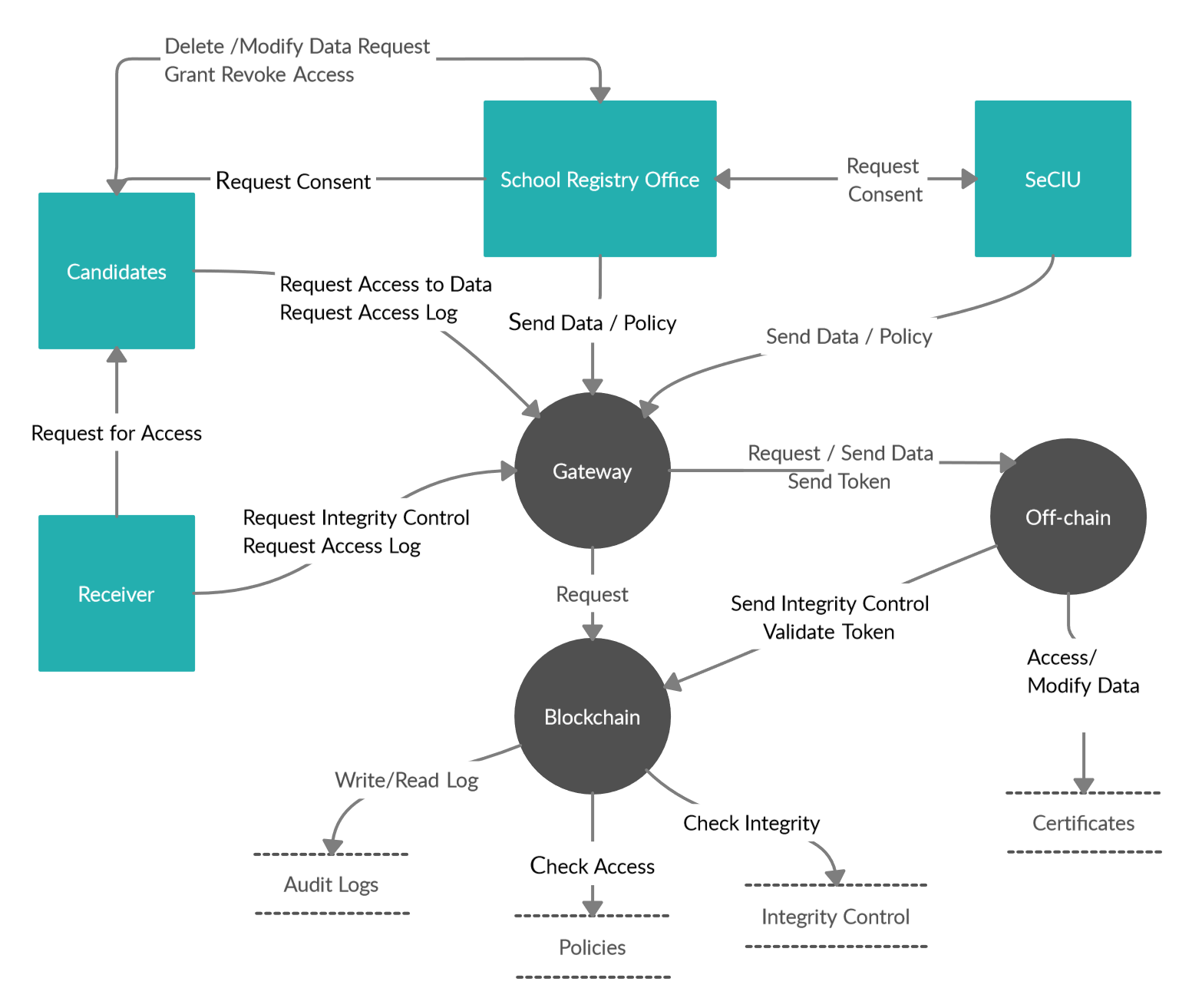}
   \caption{Data Flow Diagram of the system}
   \label{fig: DFD - System}
\end{figure*}

Fig.~\ref{fig: DFD - System} depicts the data flow diagram (DFD) of the system, detailing the data and control flow that take place. 

\subsection{Refinement of the generic use cases and diagrams}
The behavior of the digital certificates system is then specified using refined versions of the generic use cases introduced in the use case model discussed in Section~\ref{framework}.


\subsubsection{Register Personal Data} 
\begin{figure} [h]
 \centering
    \graphicspath{{./images/}}
    \includegraphics [width = 0.48 \textwidth] {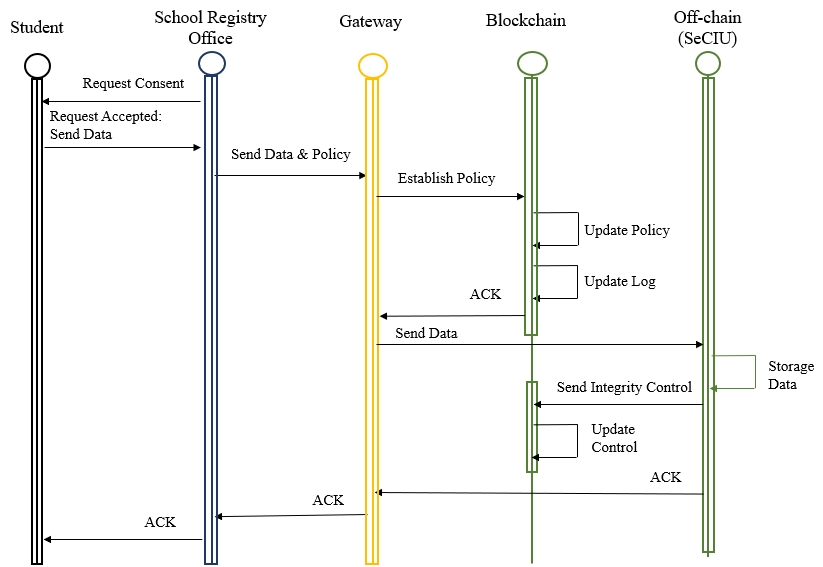}
   \caption{Sequence diagram POC: Register Personal Data}
   \label{fig: SD - SeCIU_Register Access}
\end{figure}

Registration is carried out by the School Registry Offices. Thus, they are in charge of obtaining the authorization to manage data from its corresponding owner.  The School Registry Offices send the data to the Gateway, which in turn sends the access control policy to the blockchain. In the blockchain the policy and logs are updated. Once the adjustments have been made, the registration is confirmed and the Gateway sends the student data to the off-chain, it. which stores the information and sends the integrity control information to the blockchain. Finally, the Gateway confirms to the School Registry Office that the information has been correctly stored. The sequence diagram is described in Fig.~ \ref{fig: SD - SeCIU_Register Access}.

\subsubsection{Grant Access}  
\begin{figure} [h]
 \centering
    \graphicspath{{./images/}}
    \includegraphics [width = 0.48 \textwidth] {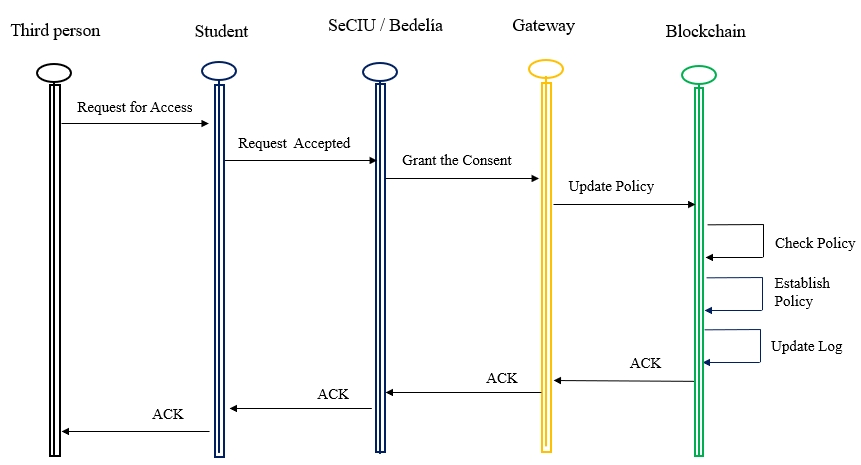}
   \caption{Sequence diagram POC: Grant Access}
   \label{fig: SD - SeCIU_Grant Access}
\end{figure}

In order for a third party (an employer or another educational institution) to either access or verify a certificate, the student must give his consent and  the system notified. This authorization will be stored in the blockchain network. The sequence diagram is described in Fig.~ \ref{fig: SD - SeCIU_Grant Access}.

\subsubsection{Revoke Access} 
\begin{figure} [h]
 \centering
    \graphicspath{{./images/}}
    \includegraphics [width = 0.48 \textwidth] {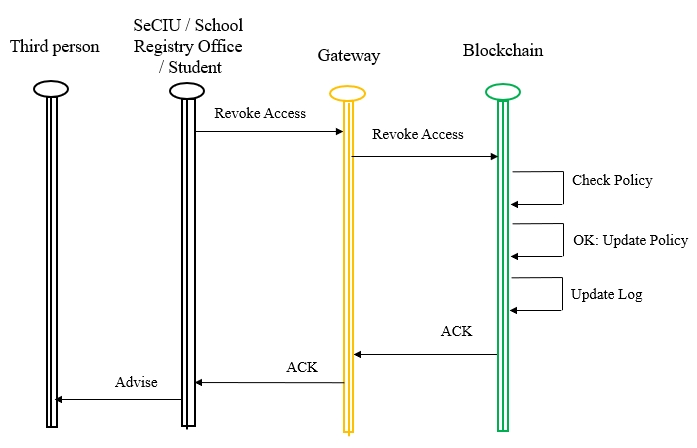}
   \caption{Sequence diagram POC: Revoke Access}
   \label{fig: SD - SeCIU_Revoke Access}
\end{figure}

As described in Fig. \ref{fig: SD - SeCIU_Revoke Access}, the access authorization given by the student can be revoked. This can be requested by the student, the School Registry Office or the SeCIU.

\subsubsection{Access to data} 
\begin{figure} [h]
 \centering
    \graphicspath{{./images/}}
    \includegraphics [width = 0.48 \textwidth] {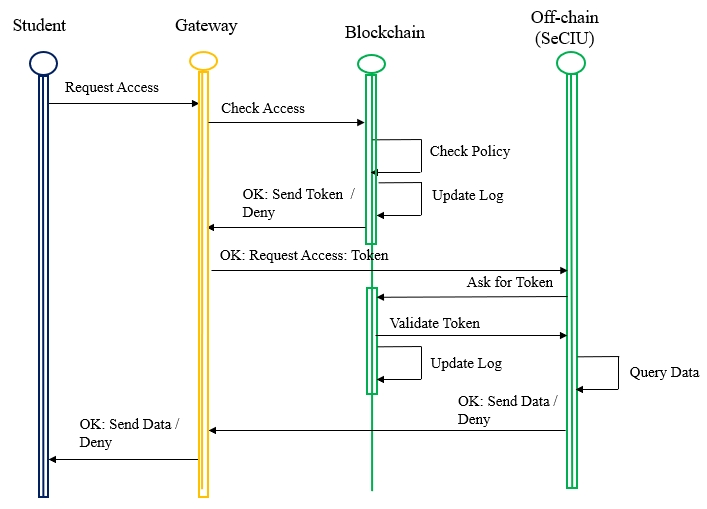}
   \caption{Sequence diagram POC: Data Access}
   \label{fig: SD - SeCIU_Data Access}
\end{figure}

In this case, described in Fig.~ \ref{fig: SD - SeCIU_Data Access}, it is considered that the access to the certificates will be carried out only by the students. The student sends a request to the Gateway, which in turn requests access to the access control blockchain network in order  to validate the access policy. In case the access is authorized, the blockchain sends to the Gateway a token. With this information the Gateway can make the request to the off-chain (where the certificate is hosted), which validates the token in the blockchain before giving the information. Then the Gateway sends the data to the student.  If a privilege granting function is implemented in the system, it would also be possible for a third party with the right privileges to access that same information.


\subsubsection{Verify data} 
\begin{figure} [h]
 \centering
    \graphicspath{{./images/}}
    \includegraphics [width = 0.48 \textwidth] {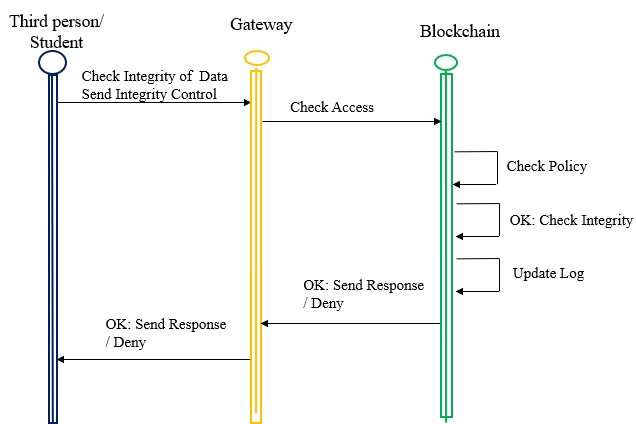}
   \caption{Sequence diagram POC: Verify Data}
   \label{fig: SD - SeCIU_Verify Data}
\end{figure}

Employers or institutions that want to verify a certificate (that could have been delivered by the student), can do so by consulting the Gateway if the student previously made the corresponding authorization. For this reason the Gateway first verifies this access, before carrying out the integrity check. The sequence diagram is described in Fig.~ \ref{fig: SD - SeCIU_Verify Data}.

\subsubsection{Delete/Modify Data from Owner} 
\begin{figure} [h]
 \centering
    \graphicspath{{./images/}}
    \includegraphics [width = 0.48 \textwidth] {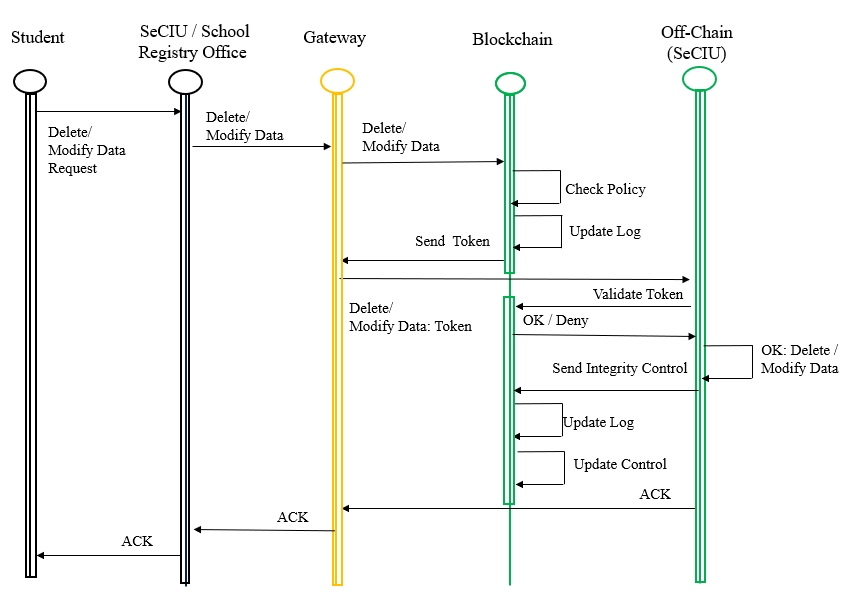}
   \caption{Sequence diagram POC: Delete Data from DO}
   \label{fig: UCD - SeCIU_Delete Data from DO}
\end{figure}

It could be the case that a student requests to modify stored information owned by him. That request  is evaluated and processed by the corresponding School Registry Office, passing the request to the gateway. The sequence diagram is described in Fig.~\ref{fig: UCD - SeCIU_Delete Data from DO}. The change involves checking the access policy, updating the information in the off-chain and updating the integrity control in the blockchain. 


\subsubsection{Delete/Modify Data from DC or DP} 
\begin{figure} [h]
 \centering
    \graphicspath{{./images/}}
    \includegraphics [width = 0.48 \textwidth] {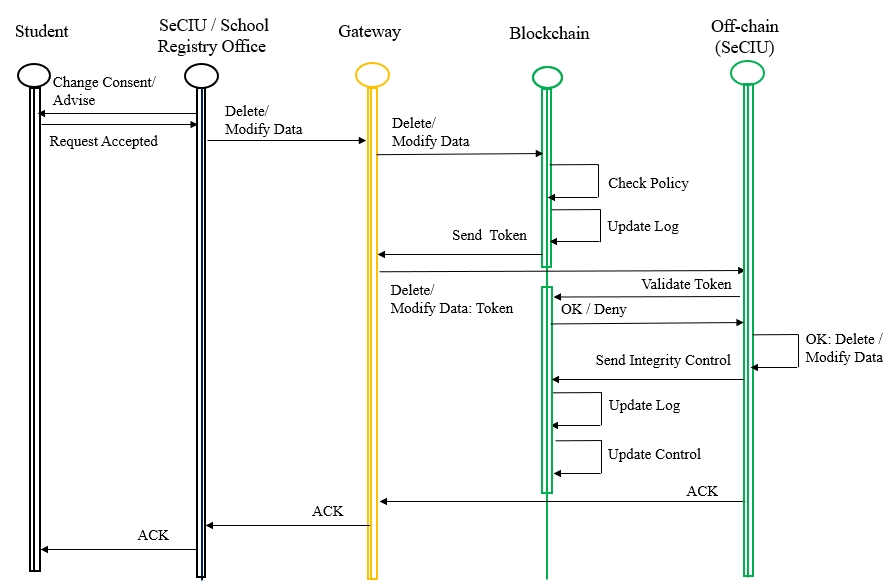}
   \caption{Sequence diagram POC: Delete Data from DC}
   \label{fig: SD - SeCIU_Delete Data from DC}
\end{figure}

Similarly, as is shown in Fig.~ \ref{fig: SD - SeCIU_Delete Data from DC}, it may be SeCIU or the School Registry Office who must modify a student's data (for example, the correction of a grade). The cycle is similar to the previous one, but notice of the operation is also given to the student.

\subsubsection{Request for Access Log} 
\begin{figure} [h]
 \centering
    \graphicspath{{./images/}}
    \includegraphics [width = 0.48 \textwidth] {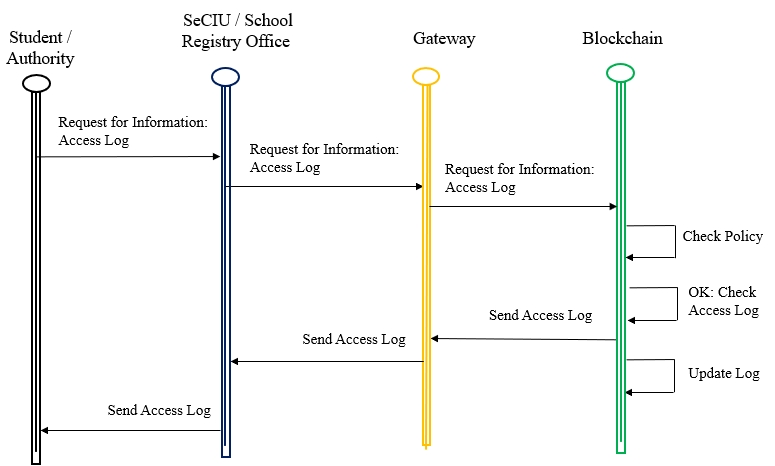}
   \caption{Sequence diagram POC: Access Log}
   \label{fig: SD - SeCIU_Access Log}
\end{figure}

Finally, the student or a supervisory authority may request audit information of the accesses to certificates. To do this the solicitor makes the request to the SeCIU or School Registry Office, who consults the Gateway. The Gateway send a request to the blockchain, who validates the request in its access policy and sends the requested log information. The sequence diagram is described in Fig.~\ref{fig: SD - SeCIU_Access Log}.

\section{Conclusion and further work} 
\label{conclusion}
We have identified and discussed the challenges that the European regulation on data protection poses to the design and implementation of software systems that manage personal data. In particular we are interested in those systems that are built using blockchain technology. 

One first and quite straightforward conclusion we make from the analysis we have developed  is that complementary mechanisms are needed for those solutions to comply with the GDPR requirements. We propose the use of off-chain mechanisms to design and implement hybrid blockchain systems with capabilities to enforce access control to personal data and to guarantee the  confidentiality  of the data managed by the system.

Having interacted in these last years with IT systems developers interested in exploiting the functionalities provided by the blockchain technology, we are convinced of the need this community has of counting with methodological tools that provide support for constructing functionally correct and privacy aware solutions, systematically. In this paper we have presented and discussed a methodological tool that contributes in that direction.

As future work it remains to complete the implementation of the prototype in order to assess the adequacy of the proposed design framework. An identified challenge is the conception of the mechanisms to register and authenticate the third parties that shall interact with the digital certificates system. In particular, as SECIU would prefer not to provide and manage those mechanisms, a decentralized solution must be considered. How well one such solution would integrate with the rest of system preserving the privacy requirements requires further study.
\newpage
\bibliographystyle{IEEEtran}
\bibliography{biblio_BCPDP}
\newpage
\appendix
\label{seqdiagrams}
In what follows we provide a complete specification of the use case model presented in Section~ \ref{framework}



\subsubsection{Register personal data use case}
\begin{figure} [h]
 \centering
    \graphicspath{{./images/}}
    \includegraphics [width = 0.45 \textwidth] {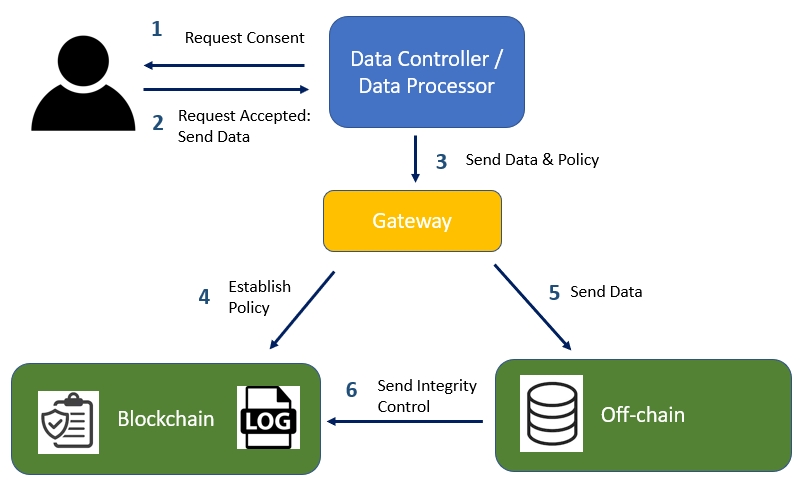}
   \caption{Use case diagram: Register Personal Data}
   \label{fig: UCD - Register Personal Data}
\end{figure}

\begin{figure} [h]
 \centering
    \graphicspath{{./images/}}
    \includegraphics [width = 0.48 \textwidth] {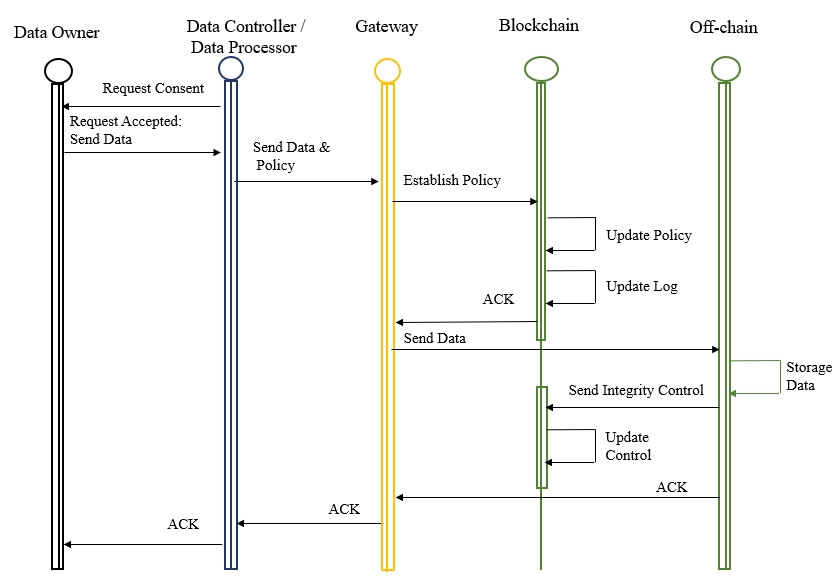}
   \caption{Sequence diagram: Register Personal Data}
   \label{fig: SD - Register Access}
\end{figure}

The first use case corresponds to entering the personal data into the system, as it is described in Fig.~\ref{fig: UCD - Register Personal Data}. In order to register personal data, either the  data controller (DC) or the data processor (DP) must first obtain authorization from the data owner (DO) to record the information and establish a policy regulating the access to that information. This can be done by submitting a request signed by the user and either the DC or the DP. Then, the data is sent to the Gateway, that validates the access intent with the access control network blockchain. If the access policy  authorizes the registration, the Gateway sends the data to the off-chain. Finally, the off-chain network communicates with the integrity network blockchain to stores an integrity control (a hash of the personal data). All these changes are stored in turn in the audit network blockchain. The sequence diagram is described in Fig.~\ref{fig: SD - Register Access}.

\subsubsection{Grant Access use case}
\begin{figure} [h]
 \centering
    \graphicspath{{./images/}}
    \includegraphics [width = 0.45 \textwidth] {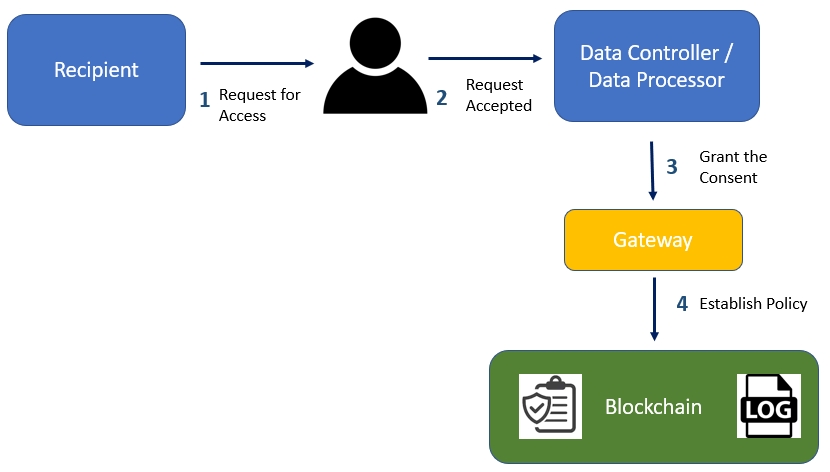}
   \caption{Use case diagram: Grant Access}
   \label{fig: UCD - Grant Access}
\end{figure}

\begin{figure} [h]
 \centering
    \graphicspath{{./images/}}
    \includegraphics [width = 0.48 \textwidth] {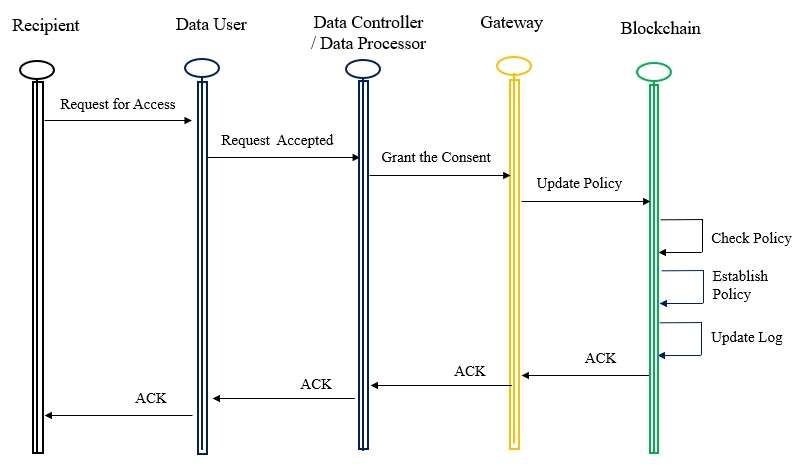}
   \caption{Sequence diagram: Grant Access}
   \label{fig: SD - Grant Access}
\end{figure}

The second flow considers the case where access to a third party (recipient) is processed. As shown in Fig.~\ref{fig: UCD - Grant Access}, in order for a recipient to access the information, the DO and the DC or DP must authorize the request and send this authorization to the blockchain, so that the access policy is updated. This implies, therefore, registering the recipient in the system. It should be noted that the off-chain does not intervene in this flow since the data is neither accessed nor modified.  The sequence diagram is described in Fig.~\ref{fig: SD - Grant Access}.

\subsubsection{Revoke Access use case} 
\begin{figure} [h]
 \centering
    \graphicspath{{./images/}}
    \includegraphics [width = 0.45 \textwidth] {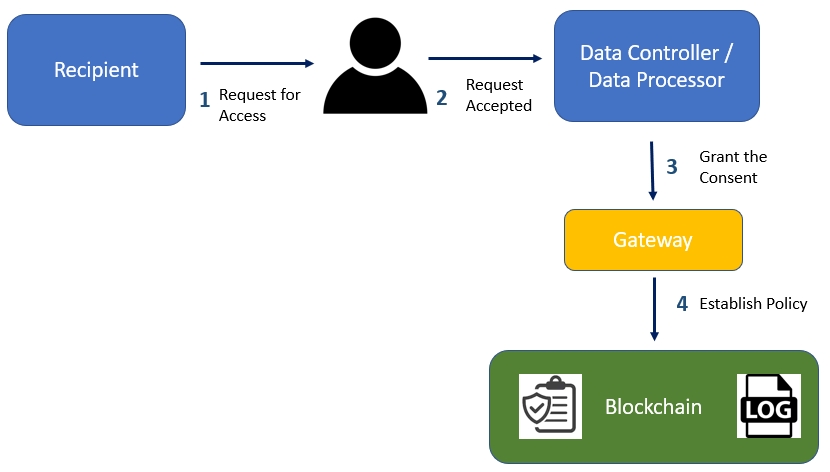}
   \caption{Use case diagram: Revoke Access}
   \label{fig: UCD - Revoke Access}
\end{figure}

\begin{figure} [h]
 \centering
    \graphicspath{{./images/}}
    \includegraphics [width = 0.48 \textwidth] {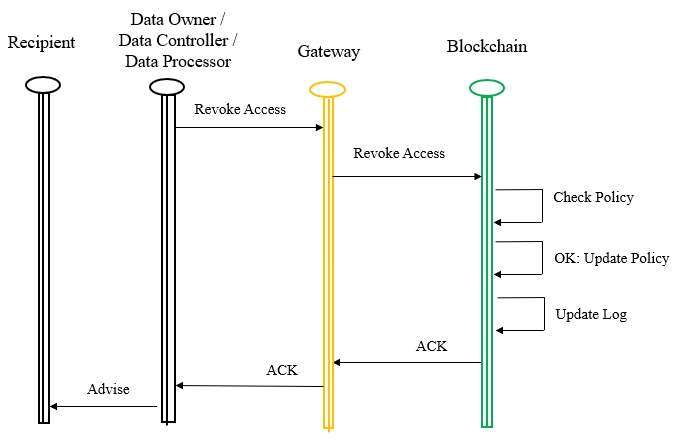}
   \caption{Sequence diagram: Revoke Access}
   \label{fig: SD - Revoke Access}
\end{figure}

The process of removing permissions involves updating the access policy stored on the blockchain. This action can be done by the DO, DC or DP. The blockchain first validates the identification and access authorization, before processing the requested change. Finally the change in the audit is recorded. The use case is represented in Fig.~\ref{fig: UCD - Revoke Access} and the sequence diagram is described in Fig.~\ref{fig: SD - Revoke Access}.

\subsubsection{Data access use case} 
\begin{figure} [h]
 \centering
    \graphicspath{{./images/}}
    \includegraphics [width = 0.35 \textwidth] {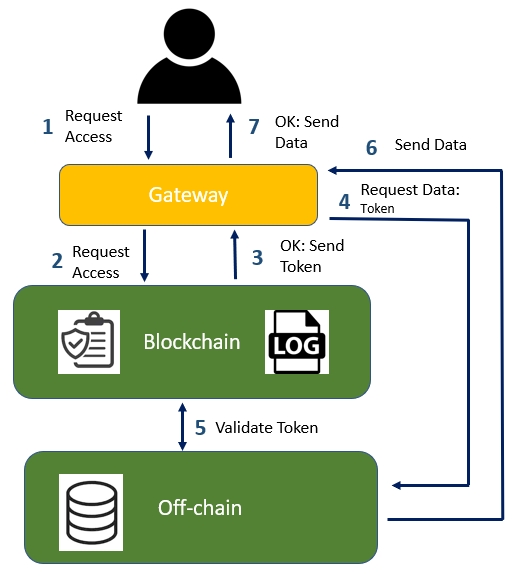}
   \caption{Use case diagram: Data Access}
   \label{fig: UCD - Data Access2}
\end{figure}

\begin{figure} [h]
 \centering
    \graphicspath{{./images/}}
    \includegraphics [width = 0.48 \textwidth] {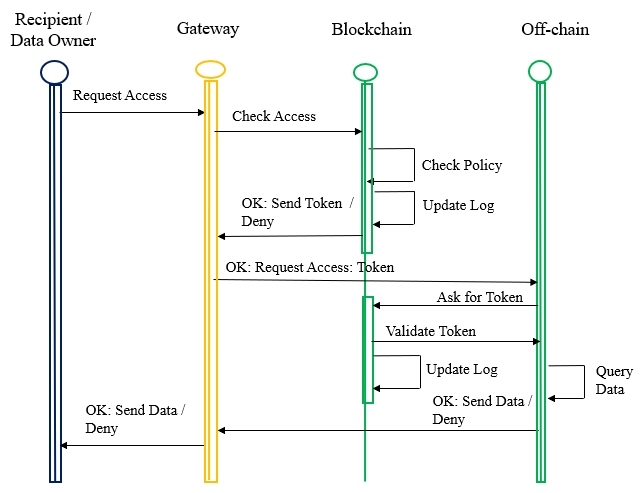}
   \caption{Sequence diagram: Data Access}
   \label{fig: SD - Data Access2}
\end{figure}

This use case has been described in Section \ref{framework}. The use case is represented in Fig.~\ref{fig: UCD - Data Access2} and the sequence diagram is described in Fig.~\ref{fig: SD - Data Access2}.

\subsubsection{Verify data use case} 
\begin{figure} [h]
 \centering
    \graphicspath{{./images/}}
    \includegraphics [width = 0.32 \textwidth] {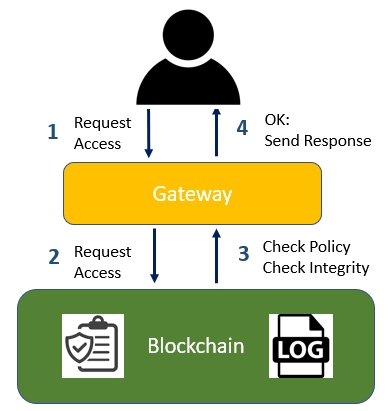}
   \caption{Use case diagram: Verify Data}
   \label{fig: UCD - Verify Data2}
\end{figure}

\begin{figure} [h]
 \centering
    \graphicspath{{./images/}}
    \includegraphics [width = 0.48 \textwidth] {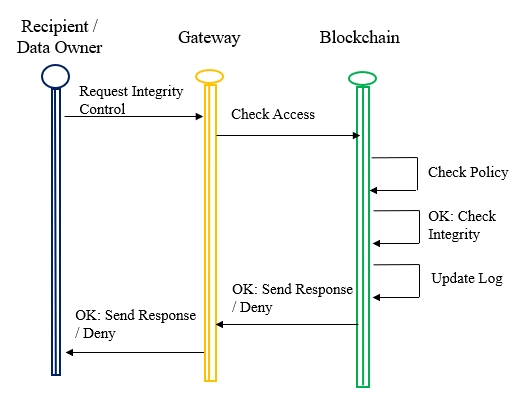}
   \caption{Sequence diagram: Verify Data}
   \label{fig: SD - Verify Data2}
\end{figure}

This use case has been described in Section \ref{framework}. The use case is represented in Fig.~\ref{fig: UCD - Verify Data2} and the sequence diagram is described in Fig.~\ref{fig: SD - Verify Data2}.

\subsubsection{Delete/Modify Data from Data Owner} 
If the DO wishes to modify or delete his information, he can communicates it to the DC or DP who send the request to the Gateway. The Gateway verifies the policy against the access control network blockchain who sends a token to validate the access (similar to the register personal data use case). The Gateway sends the request to the off-chain together with the token, so that the off-chain can verify it against the blockchain. If the token is validated, the off-chain network makes the change and adjusts the integrity check hash. The use case diagram is illustrated in Fig.~\ref{fig: UCD - Delete Data from DO} and the sequence diagram is described in Fig.~\ref{fig: UCD - Delete Data from DO}.

\begin{figure} [h]
 \centering
    \graphicspath{{./images/}}
    \includegraphics [width = 0.45 \textwidth] {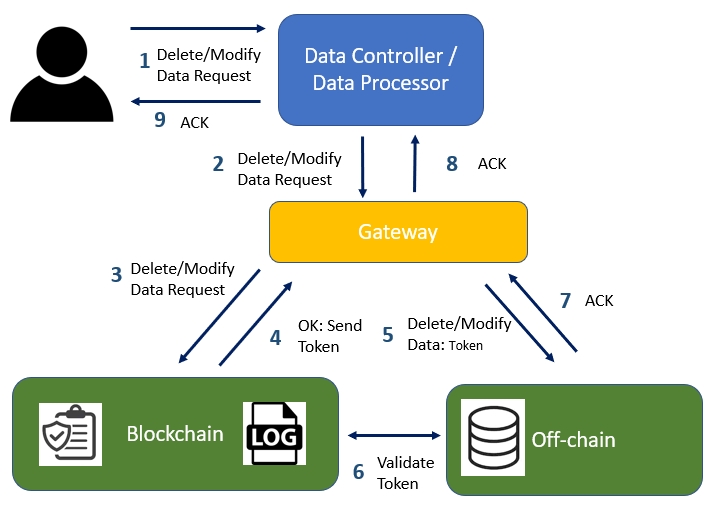}
   \caption{Use case diagram: Delete Data from DO}
   \label{fig: UCD - Delete Data from DO}
\end{figure}

\begin{figure} [h]
 \centering
    \graphicspath{{./images/}}
    \includegraphics [width = 0.48 \textwidth] {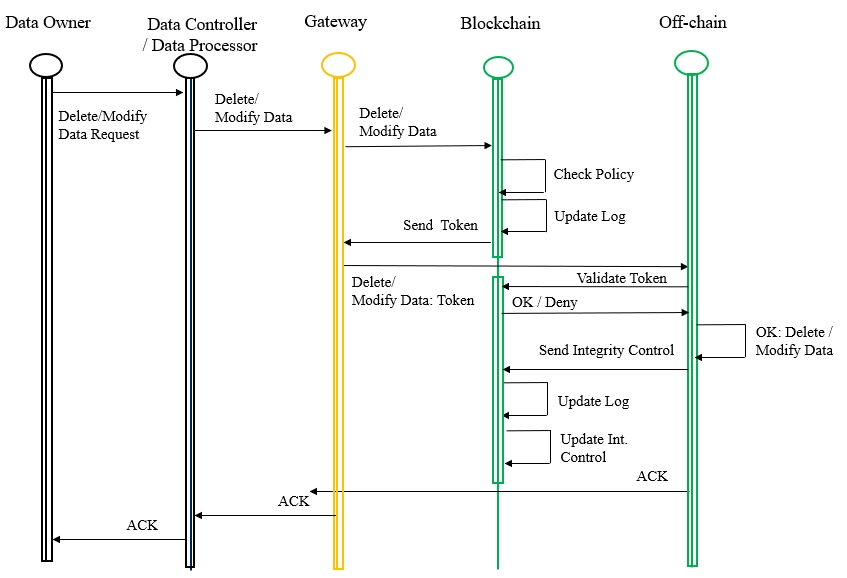}
   \caption{Sequence diagram: Delete Data from DO}
   \label{fig: UCD - Delete Data from DO}
\end{figure}


\subsubsection{Delete/Modify Data from DC or DP} 
\begin{figure} [h]
 \centering
    \graphicspath{{./images/}}
    \includegraphics [width = 0.45 \textwidth] {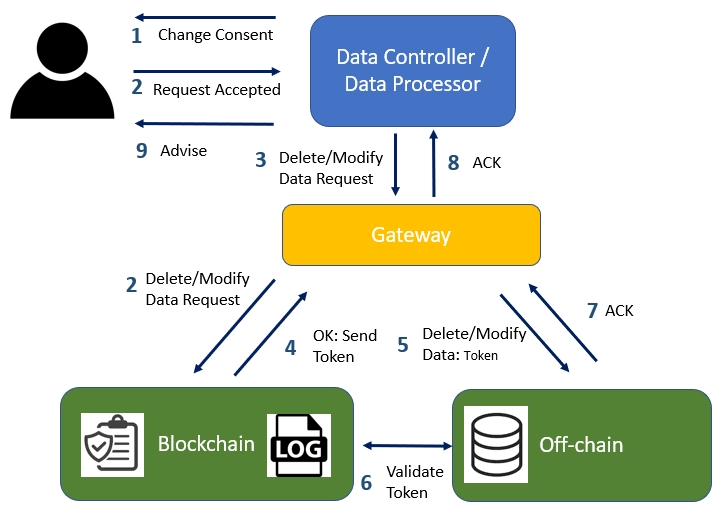}
   \caption{Use case diagram: Delete Data from DC or DP}
   \label{fig: UCD - Delete Data from DC}
\end{figure}

\begin{figure} [h]
 \centering
    \graphicspath{{./images/}}
    \includegraphics [width = 0.48 \textwidth] {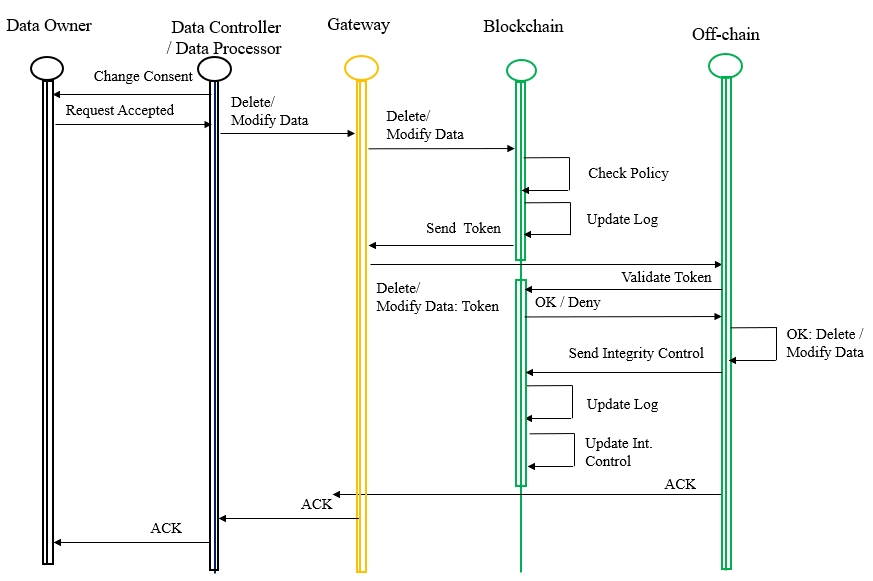}
   \caption{Sequence diagram: Delete Data from DC or DP}
   \label{fig: SD - Delete Data from DC}
\end{figure}

In turn, as is described in in Fig.~\ref{fig: UCD - Delete Data from DC} , the DC or DP can modify or delete the information they have stored, giving notice to the data owner of the setting. Depending on the action to be taken and the context, it may be necessary to previously request the authorization of the DO. For example, we understand that in the case of deletion of information, it is only required to report the fact to the data owner, while the modification of a data may require prior authorization. The sequence diagram is described in Fig. \ref{fig: SD - Delete Data from DC}.

\subsubsection{Request for Access Log} 
\begin{figure} [h]
 \centering
    \graphicspath{{./images/}}
    \includegraphics [width = 0.45 \textwidth] {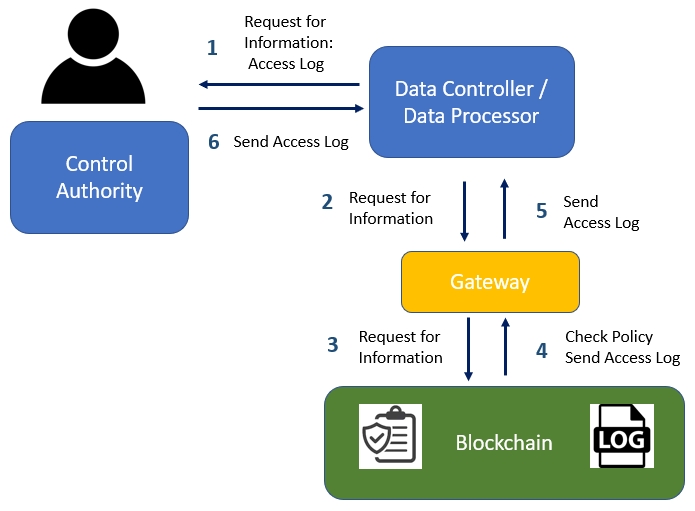}
   \caption{Use case diagram: Access Log}
   \label{fig: UCD - Access Log}
\end{figure}

\begin{figure} [h]
 \centering
    \graphicspath{{./images/}}
    \includegraphics [width = 0.48 \textwidth] {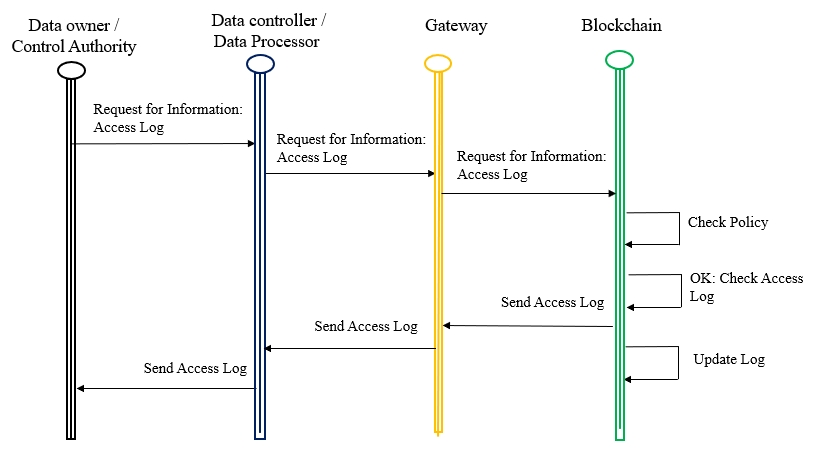}
   \caption{Sequence diagram: Access Log}
   \label{fig: SD - Access Log}
\end{figure}

Finally, the data owner or an authority can request audit information from the DC or DP, who consult it with the blockchain. As in the previous cases, before providing the information, the access policy must be reviewed. The use case diagram is illustrated in Fig.~\ref{fig: UCD - Access Log}, as the sequence diagram is described in Fig.~\ref{fig: SD - Access Log}.

\end{document}